\documentclass[traditabstract]{aa}  

\usepackage{natbib}
\bibpunct{(}{)}{;}{a}{}{,} 
\usepackage{graphicx}
\usepackage{txfonts}
\begin{document}
   \title{Radial Velocities with CRIRES\thanks{Based on observations 
   taken at the VLT (Paranal), under programs 280.C-5064(A) and 60.A-9051(A), and with the CORALIE spectrograph at the Euler Swiss telescope (La Silla).
}}

   \subtitle{Pushing precision down to 5-10 m/s}

   \author{P. Figueira\inst{1}, 
          F. Pepe\inst{1},
          C. H. F. Melo\inst{2},
          N. C. Santos\inst{3},
          C. Lovis\inst{1},
          M. Mayor\inst{1},
          D. Queloz\inst{1},
          A. Smette\inst{4},
                     \and 
           S. Udry\inst{1}
          }

   \institute{Geneva Observatory, University of Geneva, 51 Ch. des Maillettes, 
   1290-Sauverny, Switzerland\\
              \email{pedro.figueira@unige.ch}
          \and 
         	    European Southern Observatory, Karl-Schwarzschild-Strasse 2, D-85748 Garching bei Muenchen, Germany 
         \and
            Centro de Astrof\'{i}sica, Universidade do Porto, Rua das Estrelas, 4150-762 Porto, Portugal
	\and
               European Southern Observatory, Alonso de C\'{o}rdova 3107, Vitacura, Casilla 19001, Santiago 19, Chile      
          }

   \date{}

  \abstract{With the advent of high-resolution infrared spectrographs, Radial Velocity (RV) searches enter into a new domain. As of today, the most important technical question to address is which wavelength reference is the most suitable for high-precision RV measurements. 

In this work we explore the usage of atmospheric absorption features. We make use of CRIRES data on two programs and three different targets. We re-analyze the data of the TW\,Hya campaign, reaching a dispersion of about 6\,m/s on the RV standard in a time scale of roughly 1 week. We confirm the presence of a low-amplitude RV signal on TW\,Hya itself, roughly 3 times smaller than the one reported at visible wavelengths. We present RV measurements of Gl\,86 as well, showing that our approach is capable of detecting the signal induced by a planet and correctly quantifying it. 

 Our data show that CRIRES is capable of reaching a RV precision of less than 10\,m/s in a time-scale of one week. The limitations of this particular approach are discussed, and the  limiting factors on RV precision in the IR in a general way. The implications of this work on the design of future dedicated IR spectrographs are addressed as well.}


\authorrunning{P. Figueira et al.}
\titlerunning{Pushing down RV precision in the IR with CRIRES}

   \maketitle
%

\section{Introduction}

   The discovery of a Hot Jupiter orbiting 51\,Peg by \citet{1995Natur.378..355M} triggered the quest for extrasolar planets. This campaign made use of the radial velocity technique (henceforth RV), by which the presence of a planetary companion is inferred by the wobble it induces on the parent star. Several dedicated surveys have employed this method, concentrating in the beginning on solar-type stars, the most favorable target for high-precision measurements. However, soon it became clear that extrasolar planets were ubiquitous around these hosts. Longing for a better understanding of formation processes, planetary searches started to diversify themselves, targeting both young \citep{2006A&A...449..417S, 2007ApJ...660L.145S} and evolved \citep{2007A&A...472..657L, 2008PASJ...60.1317S}, low mass \citep{2004ASPC..318..286B, 2008ApJ...673.1165E} and massive stars \citep{2005A&A...437..743H, 2005A&A...444L..21G}. 
   
   Recent technological developments allowed for more precise spectrographs to be built, such as HARPS \citep{2003Msngr.114...20M}, and reduction and analysis methods perfected themselves throughout the years \citep{2007A&A...468.1115L}. As of today, HARPS yields the most accurate RV measurements, with sub-m/s precision, allowing for a succession of ground-breaking detections of the lightest planets known \citep{2006Natur.441..305L, 2009A&A...493..639M}.  
   
   Both the search for more exquisite stellar hosts and the race for the lightest planets point towards M dwarfs. These stars, the most abundant in the Universe, are also the lightest ones. Since the RV variation induced by a planet on a star scales with $M^{-2/3}$, the amplitude of the effect induced on an M star is significantly larger. As an example, a planet of identical mass at the same distance from the stellar host produces an RV variation with an amplitude $\sim$\,3 times larger on an M5 star than on a G2 star. The drawback is that M dwarfs, being much colder, are much fainter in optical wavelengths. The RV survey of light-mass stars points then towards the exploration of a new wavelength domain, the IR, where the luminosity of these objects peaks.
   
   As RV searches started to target young and active stars, another advantage of the IR became apparent. RV signals can be created by surface inhomogeneities, like stellar spots \citep{2001A&A...379..279Q}. Since the nature of these phenomena can bypass current diagnosis method such as bisector analysis \citep[see, for instance,][]{2007A&A...473..983D}, a planetary origin can be assigned to a periodic variation which is, in fact, of stellar origin. By observing in the IR, as in opposition to the visible wavelengths, one observes in a domain where the contrast between stellar spots and the stellar disk is reduced. Since the spot\,/\,star contrast ratio is at the root of the RV signal detected, the amplitude of the latter is reduced as well \citep{2006ApJ...644L..75M, 2008A&A...489L...9H, 2008ApJ...687L.103P}. By comparing RV drawn from optical and IR spectrographs, one can assign the planetary or stellar nature to a RV signal in an unambiguous way.
   
However, the exploration of the RV in the IR is in its infancy, and many problems remain to be tackled. In this work we concentrate ourselves in describing a precise wavelength calibration technique for the IR domain. In section 2 we discuss the difficulties of wavelength calibration in the IR and present the principles of our technique. In section 3 we describe CRIRES and the observations, and how we applied our approach. Section 4 presents an outline of the data reduction and analysis. Section 5 presents our results. Section 6 discusses these results and their implications on the design of spectrographs dedicated  to RV measurements in the IR. We finish by stating our conclusions in Section 7.


\section{Wavelength References in the IR}

In order to detect small-amplitude RV signals (i.e., of m/s) a simultaneous wavelength reference must be employed. In the visible wavelengths, the Th-Ar emission lamps and the I$_{2}$ absorption cell have proved to yield a precision of better than 1\,m/s \citep{2009A&A...493..639M} and 3\,m/s \citep{2009ApJ...696...75H}, respectively. Impressive as they are, the transfer of this knowledge to the IR domain is not straightforward. The density of Th-Ar lines is much lower in the IR than in the visible, making its usage difficult. On top of that, since the lamp lines do not follow the same instrumental profile (IP) variations than the science targets (they pass through different optical paths), a stable spectrograph is required. The more frequent gas-cell approach does not provide for a better solution, as no known gas creates a dense absorption forest in a wide wavelength range in the IR. Several hybrid alternatives were put forward to attempt to solve this problem, such as gas cell plus emission lamp or coupling of different gas cells. For a detailed description of the problem and proposed solutions the reader is referred to the work of \cite{2009ApJ...692.1590M}.

A different approach can be undertaken by using one of the peculiarities of the IR: the absorption of light by the Earth's atmosphere. Our atmosphere is composed mainly of molecular species, whose rotational and vibrational features correspond to near-IR wavelengths. Usually seen as a nuisance, they can be used to our advantage. In some narrow wavelength domains, the presence of well-defined, sharp, and deep lines can provide for a good local wavelength solution. While this is only possible locally, it is well suited for the current high-resolution IR spectrographs, characterized by a small simultaneous wavelength coverage, typically of $\lambda$/50-70. As a free-of-charge, always present gas cell, the usage of the atmospheric features as a wavelength reference is to be assessed before venturing into other alternatives.

However, our ``gas-cell" suffers from a particular problem: the lines it imprints on the spectra depend on the atmospheric conditions. While some of lines' properties are expected to remain constant, others, such as depth, vary at a several \% level. It is then very hard to apply the typical deconvolution technique  \citep{1996PASP..108..500B}, which assumes a detailed knowledge of the absorbing spectra to calculate the IP and correct for it. While several dedicated models have been developed \citep{2007PASP..119..228B} to address this problem, there is no evaluation of the impact of the remaining residuals on RV, and is arguably bigger than the m/s level. As a matter of fact, the only campaign that employed the deconvolution approach in the IR in the more standard way, \cite{2007ApJ...666.1198B}, reported a precision of 300\,m/s and stated that a precision down to 100\,m/s was within reach. However, this result is to be taken with caution, as it was derived for a campaign of M dwarfs, an exigent stellar host.

We decided to depart from this approach, by avoiding the overlapping between target and reference features. While the spectra of the star and of the atmosphere are simultaneous in the sense that they are obtained at the same moment, the lines are well separated spectroscopically and can be fitted independently. The center of each line is thus determined in a straightforward way (for instance, the atmospheric lines' depth variation has no impact on the measured RVs). The IP variations that affect stellar lines affect reference lines as well. If the same fitting procedures are employed on both line sets, the effect cancels out. 

Several notable experiments on the usage of atmospheric absorption features were made in the eighties, of which we highlight the results of \cite{1982A&A...114..357B}, \cite{1982ApJ...253..727S}, and \cite{1985A&A...149..357C}. Using O$_2$ lines, these authors could reach a precision of 5\,m/s. The value is made even more relevant by the fact that they used different RV determination methods (different instrumentation, different line fitting approaches, etc.), now believed to be obsolete when compared with the Cross Correlation Function \citep{1996A&AS..119..373B} or the Deconvolution method. Very recently \cite{2004MNRAS.353L...1S} used the same principle on H$_2$O lines and reached a precision of 5-10\,m/s on UVES data. This shows that the accuracy of our wavelength reference is at least of the order of 5\,m/s, and a valuable option for us.


\section{Observations}

\subsection{CRIRES}

CRIRES (CRyogenic InfraRed Echelle Spectrograph) is the new ESO IR spectrograph mounted on the Unit Telescope 1 (UT1, Antu) of the European Southern Observatory Very Large Telescope \citep{2004SPIE.5492.1218K}. The main optical elements consist of a prism acting as a pre-disperser and a 31.6 lines/mm echelle grating. The instrument provides for a resolution of up to 100 000 when used with a 0.2" slit. It operates in the range 960-5200 nm with an instantaneous wavelength coverage of $\sim$ $\lambda/50$. The spectra is imaged on a detector mosaic, composed of four Aladdin III detectors (4096 $\times$ 512 pixel) with a gap of $\sim$250 pix between the chips. As a consequence, the instantaneous wavelength coverage is much smaller than for optical cross-dispersed spectrographs, such as HARPS, HIRES or UCLES. Adaptive Optics (MACAO - Multi-Applications Curvature Adaptive optics) can be used to optimize the signal-to-noise ratio and the spatial resolution. 

\subsection{The observing set-up}

At the time of observations, CRIRES had 200 possible observing set-ups, most of which overlapped. \footnote{Since May 2009, there are 282 standard wavelength setups available. The standard wavelength setups are designed to provide a complete coverage with no contamination by adjacent orders. In addition, users can select a free wavelength setting.} In order to select the most suitable set-up for precise RV we need to decide which one carries the higher content of useful spectral information. We stress here that by "useful" we mean domains where both reference -- i.e. atmospheric -- and target -- i.e. stellar -- spectral features are not superposed. 

We used solar NSO/Kitt Peak FTS data \citep{1991aass.book.....L, 1993aps..book.....W} to simulate CRIRES spectra. To do so we convolved the spectra with an IP which we assumed as gaussian and defined as a function of resolution (fixed at R\,=\,100\,000), and re-sampled it on the detector. This procedure was applied for the Sun spectra, atmospheric spectra, and the product of the two. 
In order to access the RV uncertainty measured on a sun-like spectrum using the atmosphere as reference we calculated the final RV uncertainty as the squared sum of the estimated error on the atmospheric lines' RV and on the stellar lines' RV. It is important to stress that the objective was not to obtain an estimation of the value of attainable precision in absolute terms but in a relative way between the different settings.

To estimate the precision on each of the components two different methods were employed: 1) a simple rule of thumb that states that the inverse of RV accuracy $\epsilon \propto \mathrm{S/N}\,.\,\bar{I} . \sqrt{n_{l}} $, where S/N is the Signal-to-noise ratio of the spectra, $ \bar{I} $ is the average depth of absorption lines and $n_{l}$  the number of lines in the spectra; 2) the Quality factor described in \cite{2001A&A...374..733B} which quantifies the spectral information present on each spectra.

It is clear that the obtained precision depends heavily on the useful spectral domain. In our particular case, the superposition of telluric and stellar spectra can deem several zones as useless for RV calculation. In order to take into account this effect, we proceeded as following. We used an automated algorithm to identify stellar lines in the simulated spectra \citep{2007A&A...469..783S}. From these we selected only stellar lines which were not superimposed on atmospheric lines bigger than a fixed threshold, and vice-versa. We added a Doppler shift in the interval [-30,+30]\, km/s in order to reproduce the line shift created by Earth's movement around the Sun (for the case of a star which has an average RV of 0 relative to us). Lines that were found to be superposed  after this step were discarded as well.
We varied the detection and superposition thresholds at the level of 1-5\%, a reasonable value for high S/N spectra. As the superposition of a relatively small line on the measured one can already introduce a RV shift of several hundreds of meters per second, it was mandatory to keep the superposition threshold low.
\smallskip

At the end of this procedure two zones were identified as the most promising: 

\begin{itemize}
\item the H band, at 1.6\,$\mu m$ - The telluric absorption is dominated by CO$_{2}$ lines, while other absorbers have negligible absorption, smaller than \,1\%. The CO$_2$ spectrum is easily identified and provides for very regular wavelength anchors, with around 30 deep lines -- from 10 to 60 \%-- on several detectors. The stellar spectra provides deep lines, which, while not being very numerous, allow for an average contrast of around 30\%.
\item the K band, at 2.32\,$\mu m$ - The atmospheric absorption shows features of several molecules, the most prominent being of CH$_4$. The stellar lines depth is below 10\%, but the high number (around 50) compensates for it. This was the setting chosen by \cite{2007ApJ...666.1198B}.
\end{itemize}

\begin{figure}
\resizebox{\hsize}{!}{\includegraphics{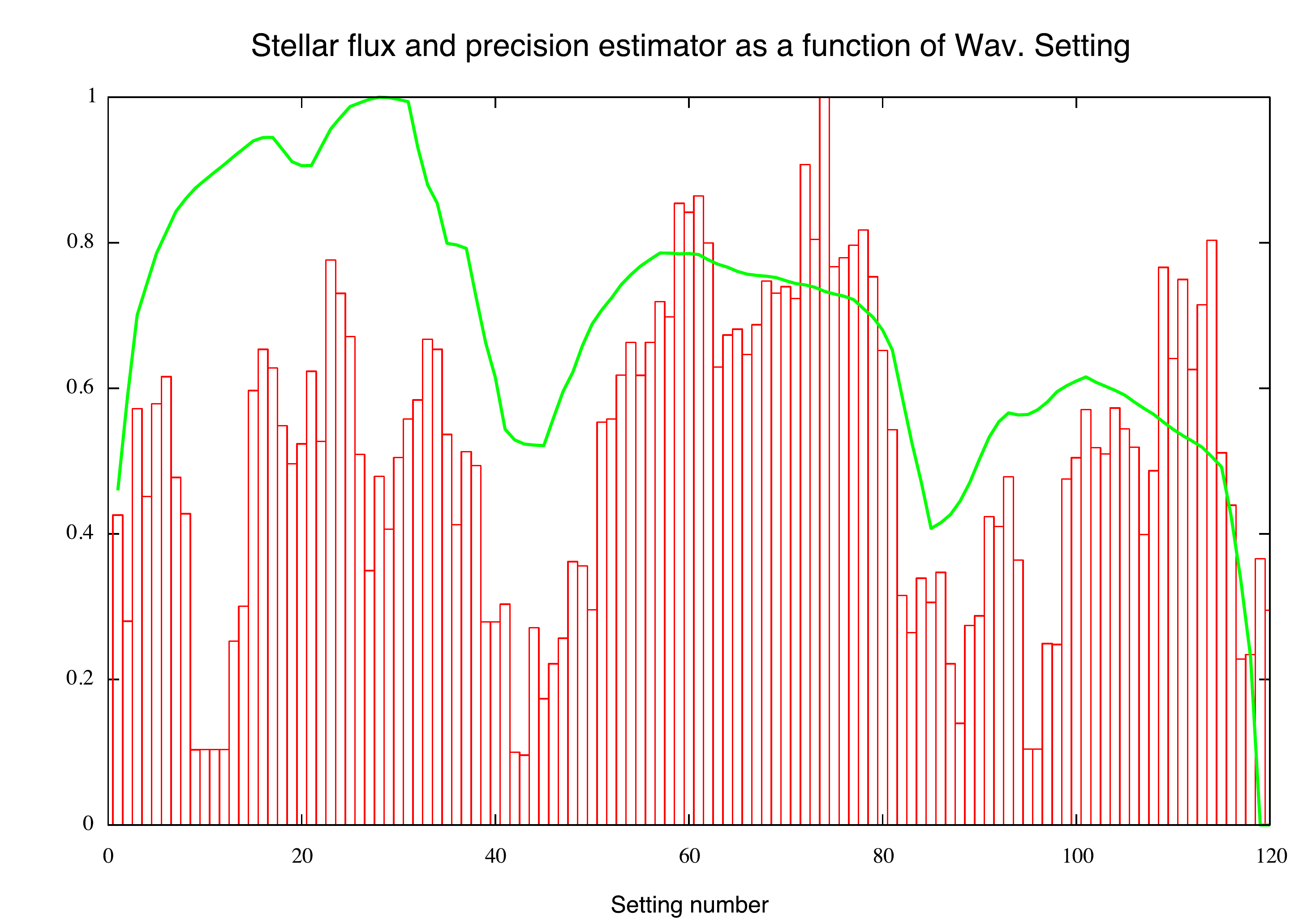}}
\caption{Precision as a function of wavelength setting, in arbitrary units, as described in Sect. 3.2 ({\it histogram, red in the electronic version}). Over-plotted is the squared root of the average solar flux, as seen by CRIRES in a given wavelength setting, in arbitrary units. ({\it thick line, green in the electronic version}). All setting short-wards of 2.4\,$\mu m$ are depicted. H band favorable zone is located between settings 60\,-\,75 and K band's around 110\,-\,115. Note that settings where numbered as a function of increasing wavelength reference, following the order presented in CRIRES manual.}\label{Fig1}
\end{figure}

We found out that the detection and superposition thresholds were the factors that decided between the two domains. For values bigger than 5\%, H band zone dominates. Figure \ref{Fig1} depicts the output of this simulation, with the spectral content increasing along the y axis. Over-plotted is the squared root of the average stellar flux, in arbitrary units as well. The final precision depends on the product of these two factors and the objective is to maximize it. Note the presence of two bumps representing the wavelength domains where atmospheric transmission becomes so low that the the precision estimators fall along with stellar flux.

The main qualitative difference between the two domains in what concerns spectral information can be summarized as: while in the K band the spectral information is extracted from more numerous and shallower atmospheric lines, in the H band it is extracted from less numerous, deeper lines. The crowding of lines in the K band increases the risk of  mutual pollution between the two sets of lines (telluric and stellar) which makes precise measurements more difficult in K band than in H band. 


\bigskip

\begin{figure}
\resizebox{\hsize}{!}{\includegraphics{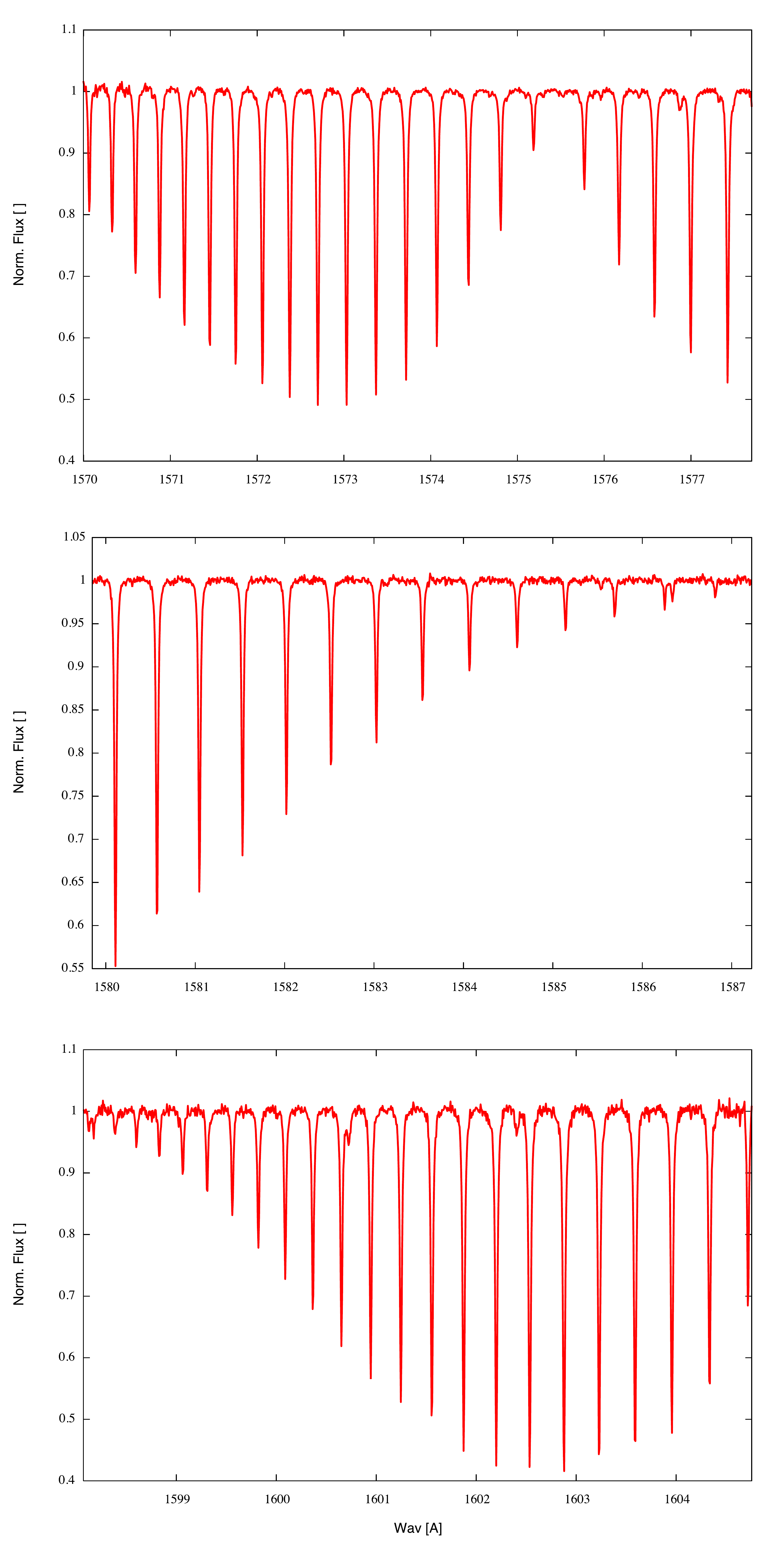}}
\caption{ Atmospheric transmission at the chosen observing setting, for the detectors 1 {\it (top)}, 2 {\it(middle)}, and 4 {\it (bottom)}.  These spectra were obtained by co-adding and then normalizing the telluric standard exposures taken during the first night.}\label{Atmtrans}
\end{figure}

The final choice was the atmospheric transmission window in the H band, and in particular the setting (36/1/i). This setting provides for deep, sharp atmospheric lines of CO$_{2}$ in detectors 1, 2 and 4. A special attention was given to image the highest number of deep stellar lines in detectors 1 and 2, as the blaze function provides for a much higher transmission in these two. The detector 1 is completely covered by a comb of 22 lines, detector 2 has 14 lines on it, and detector 4 images 20 lines. Deep lines, with relative depths of 10-50\%, provide for an homogeneous complete coverage of the first detector; in the second one the depths range  from 40 to 10\% for the first two thirds of the detector and are smaller than 10\% for the last third; finally, lines with a relative depth of 10-55\% cover 90\% of the fourth detector. It is very important to note that there are no other absorbers than CO$_2$, providing for both an unambiguous identification of spectral features and a well defined continuum around the sharp reference lines. For illustration purposes, in Fig. \ref{Atmtrans} we plot the atmospheric transmission measured by observing the telluric standard HD\,96146.

The simplified study presented here does not intend to have the last word on choice setting on CRIRES for RV studies. The optimal setting depends not only on the aforementioned superposition parameters chosen but also on the stellar spectra, the mean RV of the target and even the line identification algorithms chosen to pinpoint the lines present in each spectra. More generally than that, it depends on the RV determination method used. Our objective here is to provide for spectrally separated stellar and atmospheric lines, so that the spectral information carried by these well-defined lines can be fully exploited.

As a concluding remark on the issue we would like to note that the choice of employing the atmosphere as a wavelength reference was prompted by very practical issues. At the time of our observations there was no CRIRES gas cell available capable of providing deep absorption lines short-wards of 2.5\,$\mu m$.

\bigskip

An entrance slit of 0.2" was chosen, allowing for a resolution of $\sim$\,100 000. In order to reduce guiding errors (expected to have 1-1 translation on the lines profile and thus on the measured RV) two cautionary measures were taken. First of all, no AO was used, in order to make sure that the star PSF was spread and larger than the slit width. On top of that, observations during bad seeing periods were requested. The objective was to provide enough photons to the slit viewer/guiding camera to ensure a good centering of the star on the slit, and reduce photo-center effects on RV. While these effects affect both star and reference in the same way, minimizing their cause is the best way of ensuring a precise correspondence between the two.

\subsection{Observations}

The TW Hya campaign was performed under Director Discretionary Time (DDT), with program number 280.C-5064(A). The extensively studied TW Hya star was observed five times, from the 22$^{nd}$ to the 28$^{th}$ of February 2008, promptly followed by a RV velocity standard, HD\,108309. This star has been observed by HARPS high-precision programs and is known to be stable down to 1.1\,m/s. Between the science target and the RV standard the instrumental set-up remained fixed; no instrumental changes were allowed for. The objective was to track possible zero-point drifts of our wavelength calibration, which could have both instrumental or atmospheric origin. The observations were made using the wav. setting (36/1/i), characterized by the reference wavelength 1592.7\,nm (defined as the wavelength at the center of detector 3).

Originally, a set of 10 RV measurements of TW Hya on preferably consecutive nights was requested. However, due to scheduling difficulties aggravated by multiple instrument  interventions, the period ended with only 5 out of the 10 measurements being taken. 

\smallskip

CRIRES calibration plan comprises an RV monitoring of radial velocity standard stars with the objective of studying CRIRES' stability. From this data we recovered the observations on Gl\,86. The observations were made in the 5$^{th}$, 9$^{th}$, 17$^{th}$ and 18$^{th}$ of January 2008. The observations were made with a slit of 0.3", in the setting (36/1/n), and no AO was used. 

Table \ref{obs_table} summarizes the properties of the observations.

\begin{table*}
\centering
\begin{tabular}{lcccccc} \hline\hline
 \ \ Object &  DIT  [s] & $\#$ of Nodding pairs & Reference Wavelength [nm]  & slit width ["] & Set-up comments \\ \hline \hline
RV std   & 45   & 2 & 1592.7 & 0.2  & No AO \& No Gas-Cell  \\
TW Hya  & 180  & 2 & 1592.7 & 0.2 & No AO \& No Gas-Cell  \\
Gl 86  & 30  & 3 & 1588.4 & 0.3 & No AO \& No Gas-Cell  \\
\hline \hline
\end{tabular}

\caption{The summary of the observation details on the RV std, TW Hya, and Gl 86. All NDIT were set to 1.}\label{obs_table}
\end{table*}


\section{Data Reduction \& Analysis}

Data were reduced using a custom pipeline written in IRAF's CL\footnote{IRAF is distributed by the National Optical Astronomy Observatories, which are operated by the Association of Universities for Research in Astronomy, Inc., under cooperative agreement with the National Science Foundation.} \citep{1993ASPC...52..173T}. The procedures were tailor suited for CRIRES spectra using first the science verification data and later the data sets analyzed in this paper. It provides for automated dark and non-linearity corrections (using the nonlinearity coefficients provided by ESO), as well as flagging and replacement of bad pixels. The images are corrected from sensitivity variations by dividing by a flat-fielding corrected from blaze function effect. The nodding pairs are mutually subtracted and the orders' tracing is fitted using cubic splines. An optimal extraction algorithm is employed for extracting the spectra \citep{1986PASP...98..609H}. The objective of creating a dedicated pipeline is not only to optimize IRAF tunable knobs to provide for an optimized reduction but also to ensure a consistent and homogeneous reduction for all spectra.

The computation of RV from extracted spectra was performed using a Python and FORTRAN pipeline heavily based on HARPS' DRS (Data Reduction Software). The CO$_2$ lines' central wavelength is provided by HITRAN database \citep{2005JQSRT..96..139R}, with an accuracy of 5\,-\,50 m/s. The high S/N of the spectra and the wide separation between spectral lines allow us to fit telluric lines easily and determine their centers. For each frame, the wavelength solution was obtained by fitting by a third degree polynomial that matches the fitted centers with the theoretical wavelengths. Earth's RV in its orbit around the Sun is calculated using the ephemerides of \cite{1988A&A...202..309B}, which allows the calculus of barycentric julian dates as well. By using the Earth's RV one can transpose the wavelength solution to the Solar System barycenter. RVs are then calculated by cross-correlating the spectra against a template mask as described in \cite{1996A&AS..119..373B}. In our case we use weighted masks \citep{2002A&A...388..632P} to take into account the impact of photon noise on lines of different depth. We created two sets of masks: one from PHOENIX stellar atmosphere models \citep{2005ApJ...632.1132B} and other by measuring the position and depth of each stellar line directly on the co-added reduced spectra. The center of the cross-correlation function is determined by fitting a Gaussian function. We note that we have then the same line fit for both our reference and our science target lines. IP variations, which affect the two line sets in the same way, are fitted by the same procedure and its impact on RVs cancels out.

As described in Sec. 2, extreme caution must be taken in order to reduce mutual pollution between stellar and atmospheric spectra to a minimum. By knowing earth's RV and the approximate RV of the target\footnote{Note that the RV variations induced by putative planets are of the order of 100\,m/s and thus much smaller than the line's FWHM. As a consequence, these small amplitude fluctuations can be ignored at this stage.}, one can calculate the approximate position of each telluric and stellar line and discard those that superpose themselves. While this might lead to a significant loss of lines, it greatly reduces the systematic error introduced by blended lines. As a result the photon noise calculations on the remaining lines provide for a realistic assessment of the spectral information carried and achievable precision.

It is important to note that the RV is computed for each nodding position spectrum and detector. The RV determination is thus independent for each frame, as the always present reference lines are fitted independently, and their centers are used in the construction of a dedicated wavelength solution. The dispersion of the RV within the nodding cycle allows then for the direct calculation of the uncertainty on the averaged value for the nodding cycle. The error bar is simply the r.m.s. of the independent nodding measurements divided by the square root of the number of measurements. 

Along with each RV calculation, the bisector inverse slope was computed on the stellar cross-correlation function following the procedure described in \cite{2001A&A...379..279Q}. 


\section{Results}

\begin{figure}
\resizebox{\hsize}{!}{\includegraphics{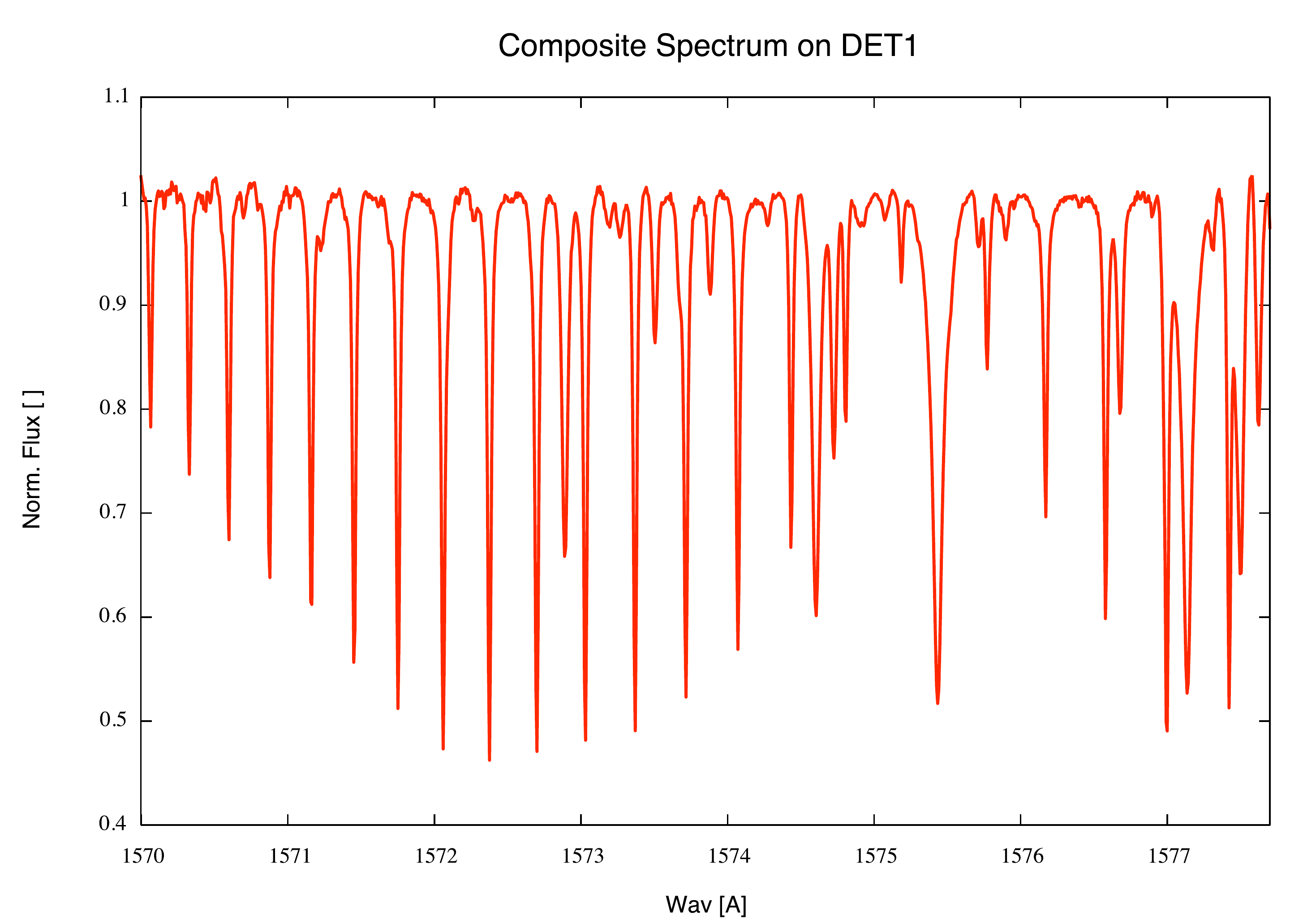}}
\caption{Spectrum of RV standard, HD\,108309, as seen on detector 1. These spectra are the result of the stacking and normalization of the 4 nodding spectra obtained during the first night. }\label{stellspec}
\end{figure}

Due to the very steep blaze function, the low S/N of TW\,Hya spectra collected in detector 4 made atmospheric line fitting impossible. On the other hand, the low number of unblended stellar lines in detector 2 combined with the lower count number led to a very high photon noise on the CCF function. For the sake of consistence and to provide for a simple and robust comparison, the standard RV presented were calculated only using the first detector as well. In Fig. \ref{stellspec} we plot the spectra of  HD\,108309, on detector 1, for illustration purposes. Out of the 22 lines, 14 unblended lines were used to calculate the wavelength solution; these lines have an average depth of 35.1$\pm$12.8$\%$ and FWHM = 3.11$\pm$0.31\,pxl.

All RV values presented here were obtained with a template mask built from the reduced CRIRES spectra. Results derived from masks built based on PHOENIX models yielded a lower precision. The dispersion of the RV standard measurements, the most reliable precision indicator among our measurements, is of $\sim$25\,m/s for PHOENIX masks. The reason for this difference is not known. 

\begin{figure*}
\includegraphics[width=17cm]{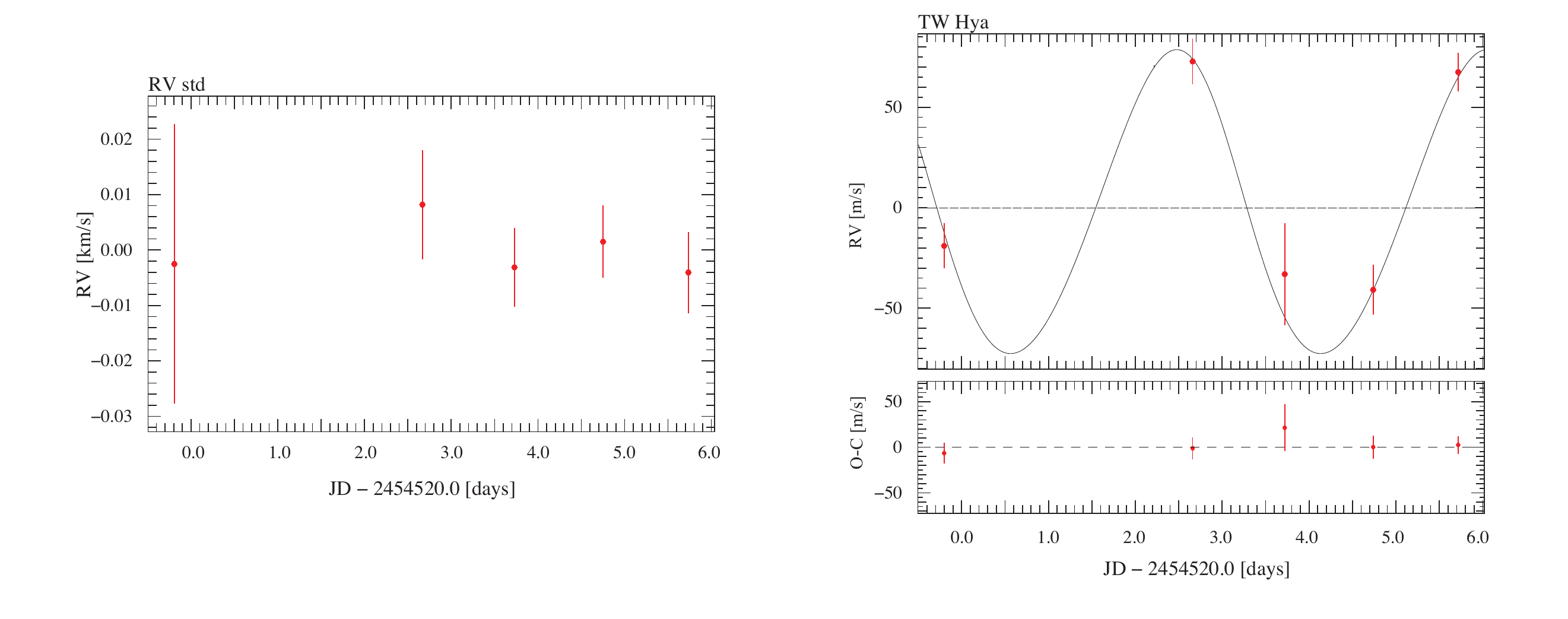}
\caption{RV variation of the standard star ({\it left}) and TW\,Hya ({\it right}) as a function of time. The error bars are drawn from the dispersion of the N=4 nodding positions, divided by 2 ($\sqrt{N}$), as described in Section 4. The optical orbit, with a different K (as described in Section 5) is over-plotted on the points.}\label{Fig2}
\end{figure*}

The RV measurements of TW\,Hya and the associated standard star are presented in Tab. \ref{RVpoints} and plotted in Fig. \ref{Fig2}.  The dispersion of the standards' RV measurements is compatible with the error bars, while for TW Hya that is not the case. Simultaneous optical RV determination were performed by CORALIE to back CRIRES observations, already described in detail in \cite{2008A&A...489L...9H}. TW Hya IR RVs are not compatible with the orbit drawn from optical RV measurements, with an amplitude of roughly 200\,m/s, defined using CORALIE data and reproducing perfectly \cite{2008Natur.451...38S} data.  To test the hypothesis of a smaller RV semi-amplitude in the IR we let this quantity and the systemic velocity as free parameters, fixing all others. The fit delivers a semi-amplitude of 80.50$\pm$6.83\,m/s with a r.m.s.(O-C) around the fit of 7.93\,m/s. The fit is over-plotted on the data. 

\begin{table*}
\centering
\begin{tabular}{lccclccc} \hline\hline
 \ \ {\bf RVstd} &MJD  [d] & RV [km/s] & Ph. Noise [m/s] & {\bf TW\,Hya} &MJD  [d] & RV [km/s] & Ph. Noise [m/s] \\ \hline \hline
    &    2454519.805826   &  30.3699   &   10.5    &   &     2454519.795056   &   12.1094   &   20.6\\
    &    2454519.806647   &  30.3600   &   10.1    &   &     2454519.797602   &   12.0837   &   19.1\\
    &    2454519.807307   &  30.3452   &   10.1    &   &     2454519.799998   &   12.1662   &   19.4\\
    &    2454519.808129   &  30.2390   &   10.7    &   &     2454519.802579   &   12.0965   &   18.9\\
    &    2454522.666534   &  30.3146   &   16.8    &   &     2454522.655881   &   12.2158   &   32.9\\
    &    2454522.667368   &  30.3778   &   16.4    &   &     2454522.658416   &   12.1857   &   26.0\\
    &    2454522.668039   &  30.3330   &   16.2    &   &     2454522.660823   &   12.2122   &   29.3\\
    &    2454522.668861   &  30.3402   &   16.7    &   &     2454522.663370   &   12.2435   &   29.9\\
    &    2454523.728265   &  30.2983   &   14.4    &   &     2454523.716034   &   12.0599   &   26.1\\
    &    2454523.729098   &  30.3300   &   14.7    &   &     2454523.718569   &   12.0379   &   23.8\\
    &    2454523.729770   &  30.3419   &   16.4    &   &     2454523.720965   &   12.1981   &   25.9\\
    &    2454523.730592   &  30.3281   &   15.7    &   &     2454523.723534   &   12.0555   &   25.8\\
    &    2454524.749738   &  30.3288   &   13.4    &   &     2454524.734402   &   12.0436   &   24.6\\
    &    2454524.750571   &  30.3301   &   12.3    &   &     2454524.736948   &   12.1082   &   22.8\\
    &    2454524.751242   &  30.3526   &   11.6    &   &     2454524.739344   &   12.0788   &   34.0\\
    &    2454524.752064   &  30.3389   &   12.5    &   &     2454524.742249   &   12.0935   &   26.1\\
    &    2454525.733788   &  30.3323   &   11.5    &   &     2454525.713207   &   12.2017   &   22.2\\
    &    2454525.734610   &  30.3690   &   10.0    &   &     2454525.715765   &   12.1959   &   18.0\\
    &    2454525.735270   &  30.3403   &   9.6    &   &     2454525.718161   &   12.1969   &   19.3\\
    &    2454525.736092   &  30.3208   &   9.8    &   &     2454525.720730   &   12.1501   &   19.7\\
\hline \hline

\end{tabular}
\caption{The RV calculated per nodding position for the RV standard, HD\,108309 and TW\,Hya, and estimated photon noise of each exposure.}\label{RVpoints}
\end{table*}

\smallskip

\begin{figure}
\resizebox{\hsize}{!}{\includegraphics{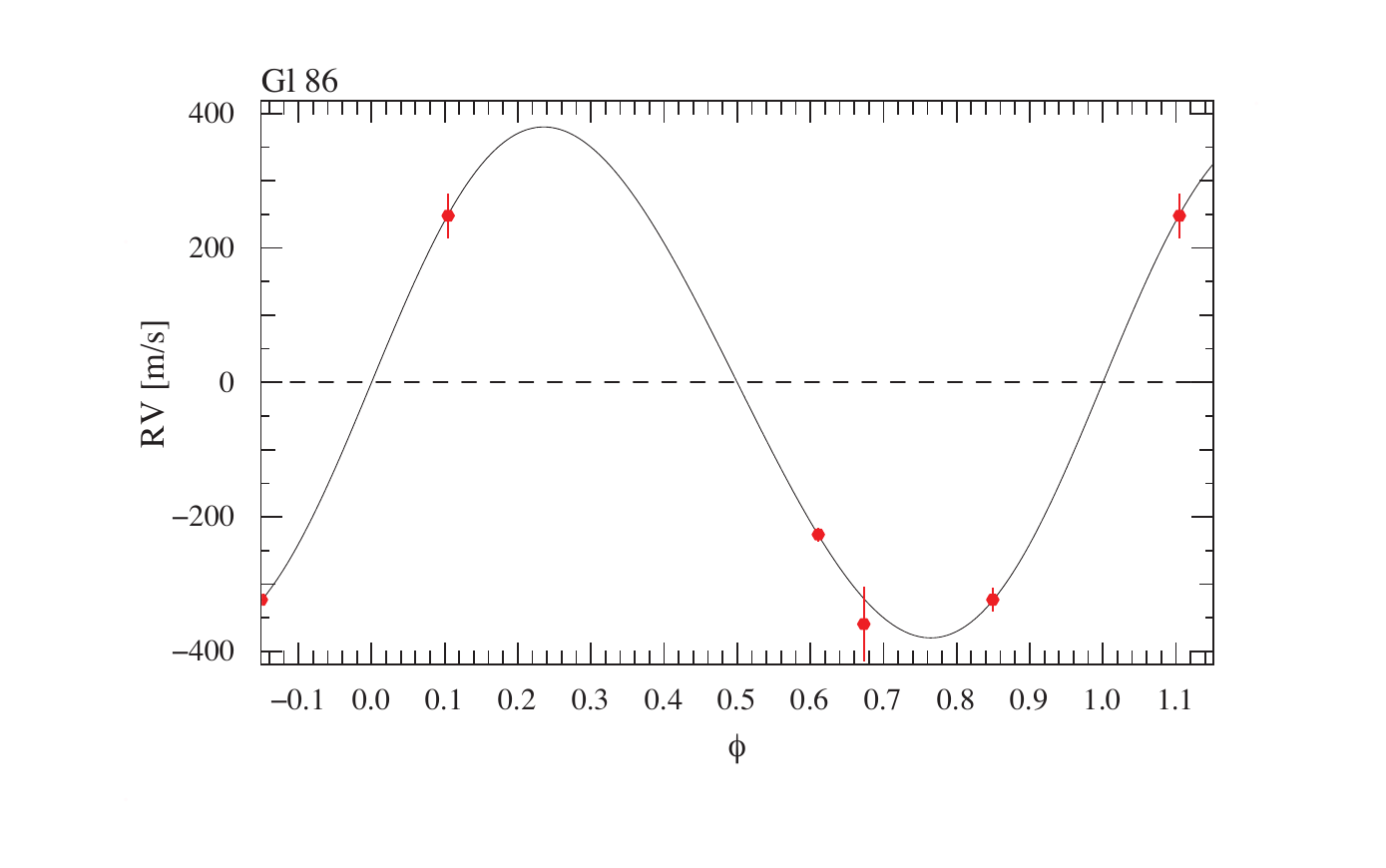}}
\caption{Phase-Folded plot of Gl\,86 Rv measurements. The error bars presented are drawn from the dispersion of the six nodding positions, divided by $\sqrt{6}$, as discussed in Section 4.}\label{Fig3}
\end{figure}

The RV measurements of Gl\,86 are presented in Tab. \ref{RVpoints_Gl86} and Fig \ref{Fig3}. The orbit presented in \cite{2000A&A...354...99Q} is superposed on the data. At first, only the systemic velocity and the T$_0$ were allowed to vary. The reason for letting T$_0$ as a free parameters comes from the fact that the error propagation from the 2000 orbit leads to a very large error on this parameter, as of 2008. The uncertainty adds to almost 8 days, which is roughly half of the orbital period. The agreement between the published curve and CRIRES measurements is shown by the small dispersion around it, of only\,5.4 m/s.
Alternatively, one can fix T$_0$ and let P to vary; the P remains unchanged within error bars, and the dispersion of the residuals remains roughly the same.
We recently obtained several CORALIE \citep{1996A&AS..119..373B} measurements on this star which, together with the 2000 data, allowed us to determine the T$_0$ in an independent way. This procedure delivered a value compatible, within error bars, with the one provided by CRIRES points alone. It is important to note that there is a measurable drift due to the presence of an outer companion \citep{2003ASPC..294...43E} and the determination of the T$_0$ at the moment of CRIRES data (when there are no CORALIE points) is affected by the shape of the unknown orbit. In any case, this T$_0$ value can be used to fit an orbit in the CRIRES data, leaving only the systemic velocity as the free parameter. This delivers a dispersion of 16\,m/s, which, while being bigger, shows the agreement between the orbit and CRIRES data.

\begin{table}
\centering
\begin{tabular}{lccc} \hline\hline
 \ \ {\bf Gl\,86} &MJD  [d] & RV [km/s] & Ph. Noise [m/s] \\ \hline \hline
    &    2454470.519076   &  55.6681   &   18.1\\
    &    2454470.519713   &  55.5957   &   16.7\\
    &    2454470.520199   &  55.6705   &   18.3\\
    &    2454470.520847   &  55.6617   &   20.4\\
    &    2454470.521345   &  55.6703   &   17.6\\
    &    2454470.521969   &  55.7322   &   17.8\\
    &    2454474.547247   &  56.3486   &   8.7\\
    &    2454474.547930   &  56.2396   &   8.4\\
    &    2454474.548508   &  56.2105   &   8.7\\
    &    2454474.549145   &  56.1806   &   8.5\\
    &    2454474.551020   &  56.1226   &   8.5\\
    &    2454474.551656   &  56.3237   &   8.1\\
    &    2454482.535667   &  55.7555   &   12.9\\
    &    2454482.536315   &  55.7780   &   11.6\\
    &    2454482.536813   &  55.7990   &   11.9\\
    &    2454482.537461   &  55.7627   &   14.3\\
    &    2454482.537970   &  55.7263   &   13.6\\
    &    2454482.538618   &  55.7589   &   12.0\\
    &    2454483.517437   &  55.8188   &   11.9\\
    &    2454483.518085   &  55.5305   &   11.8\\
    &    2454483.518571   &  55.6780   &   12.2\\
    &    2454483.519207   &  55.7346   &   12.2\\
    &    2454483.519694   &  55.4173   &   13.3\\
    &    2454483.520342   &  55.6021   &   11.5\\
\hline \hline

\end{tabular}
\caption{The RV per nodding position for Gl\,86, and estimated photon noise associated.}\label{RVpoints_Gl86}
\end{table}

\bigskip

We can estimate the precision expected on computing the cross-correlation function on each spectra, drawn from photon noise alone.
To do so we use the method presented in \cite{2001A&A...374..733B}. We considered the particular case of a gaussian absorption line and applied it to the stellar lines present in the correlation mask (the unblended ones actually used in the correlation, and of interest to us). This yields the precision on each individual spectra, i.e., on each nodding position. To calculate the error on the averaged night value we need to divide it by the square root of the number of independent nodding positions.

We can then compare three different estimators of the reached precision: 

\begin{itemize}
\item the weighted dispersion of the night-averaged RV points -- External Dispersion;
\item the error on the averaged night value, given by dispersion on the nodding cycle RV values divided by $\sqrt{N}$ -- Intra-Night Dispersion;
\item the average photon error estimator value (as described above) divided by $\sqrt{N}$ -- Photon Noise.
\end{itemize}

Table \ref{error_table} presents all these results. Note that these errors were estimated in such a way that they can be easily compared, as they provide for alternative ways of looking at the same quantity. In the cases of TW\,Hya and Gl\,86, these can be compared with the dispersion around the fit.

\begin{table*}
\centering
\begin{tabular}{lccccc} \hline\hline
 \ \ &  External Dispersion  [m/s] & Intra-Night Dispersion [m/s] & Photon Noise [m/s] & (O--C) [m/s]  \\ \hline \hline
RV std   & 5.77   & 7.03  & 6.48  & ---  \\
TW Hya  & 54.57  & 12.12  & 12.10 & 7.93  \\
Gl 86  & 122.47  & 12.77  & 7.62 & 5.41 \\
\hline \hline

\end{tabular}
\caption{The different RV precision indicators on the RV std, TW Hya, and Gl 86. The dispersion around the described fits are presented for TW\,Hya and Gl\,86.}\label{error_table}
\end{table*}


\section{Discussion}

In our previous work, \citep{2008A&A...489L...9H}, we reached a precision of 35\,m/s over a time span of roughly 1 week. This value was obtained in relative RV: science target - RV standard. This procedure allowed for a correction of the systematics still present in the data after our first reduction.  \cite{2008A&A...491..929S}, a contemporary study, reached a precision of 20\,m/s on one night. 
A dedicated effort was put in order to track systematics and understand their possible causes. As a result we could now improve the analysis procedure in order to extract the RV in an optimized way, leading to a significant increase in precision. We have no need to calculate relative RV in order to correct for systematics anymore. The RV standard dispersion is now close to photon noise, showing that there are no systematics present down to 5-10 m/s. The same is valid for TW\,Hya and Gl\,86 (O--C) residuals which show a similar dispersion. 

The RV standard displays a variation of less than 6 m/s (r.m.s.) over 1 week. The probability that a dispersion of 5.77\,m/s or lower is obtained on 5 points belonging to a distribution with $\sigma$=6.48\,m/s (the estimated photon noise) is of 58.2\%. If we consider 5 points drawn from a distribution characterized by $\sigma$=7.66\,m/s (the measured internal dispersion), the probability drops then to 40.8\%, but it is still high. We calculated the probability of drawing a dispersion of 5.77\,m/s from distributions with $\sigma$ from 5 to 10\,m/s. The results, presented in Table \ref{probs}, show that we cannot exclude that the intrinsic dispersion of our distribution is slightly larger than the measured one. As a result we consider the attained precision, in a conservative way, as being of 6-8\,m/s. 

\begin{table}
\centering
\begin{tabular}{cc} \hline\hline
 \ \ $\sigma$ [m/s] &  P  [\%]  \\ \hline \hline
5   & 84.1  \\
6  & 66.3  \\
7  & 49.9   \\
8  & 36.6   \\
9  & 26.6   \\
10  & 19.5   \\
\hline \hline

\end{tabular}
\caption{The probability of obtaining a dispersion of 5.77\,m/s on a set of 5 points characterized by dispersions in the range 5-10\,m/s.}\label{probs}
\end{table}

It is interesting to compare the values of intra-night dispersion with photon noise estimations, for the three targets. When measuring the intra-night dispersion, one takes into account not only the photon noise but also the effect of systematic errors on RV. A straightforward example of such are the errors induced by nodding and locating the target at different positions in the slit, known to have two slit jaws that are not perfectly parallel. While the values are very similar for the RV standard and TW Hya, that is not the case for Gl\,86. A possible cause is the usage of a larger entrance slit, and the consequent introduction of additional sources of error, such as stellar photo-center variations.

The RV values for the standard and TW\,Hya are compatible, within error bars, with the previously published values. This is mostly due to the comparatively large dispersion announced previously (35\,m/s) and error bars with even larger amplitude (30-60\,m/s). However, now, given the higher precision of the measurements, a low-amplitude signal emerges from the noise. As previously discussed in \citep{2008A&A...489L...9H}, a lower-amplitude signal in the IR than in the optical is expected if a spot is responsible for the RV variations instead of a planet. The previously presented tests stand as well. One can calculate the orbit parameters using optical observations. The probability that the IR RV points are drawn from this orbit by adding an error comparable to the instrumental precision is below 1$\times$10$^{-3}$\%, showing that the spot theory prevails.

The RV determinations for Gl\,86 match the orbit as established by the announcement paper. It is important to stress that for these fit the semi-amplitude, P, $e$ and $\omega$ were fixed on the published values in \cite{2000A&A...354...99Q}.  This shows that our method yields the correct value for RV signals, reproducing the {\it bona-fide} planet detected at optical wavelengths.

Given the very small number of points for both TW\,Hya and Gl\,86, the dispersion around the respective orbits is to be taken with caution. At this level the number of free parameters approaches the number of points and the noise contribution is fitted as well, leading to a $\chi^2<1$. As a representation of this, one can use the TW\,Hya data in a slightly different way. One may use the RV values obtained for each nodding position, and assign as error bars for each given by the photon noise estimation. The fit delivers, within error bars, the same orbit parameters (K and $\gamma$), a dispersion of the residuals of 35 m/s and a $\chi^2$=1.63, showing that the error bars are underestimated by photon noise alone.
At this point our objective is merely to show that the RV points are drawn from the orbits. They do not define the orbits and the dispersion around the same is not representative of the true errors for such a low number of points.

A question that arises naturally is "What are the factors limiting the RV precision delivered by this method?". This is difficult to evaluate mostly because the data set we have in hands is small. However, on theoretical ground, some educated guesses can be made. The precision on each of the telluric lines central wavelength is not quantified on HITRAN database, from which the values were drawn. The lines are merely flagged as having a precision between 5 and 50 m/s. It is thus not surprising that a precision of 5 m/s can be reached in a whole, but this might become the limiting factor if one pushes the S/N to much higher values. The presence of a systematic wavelength assignment error can also be a trivial source of error. On the other hand, atmospheric phenomena such as localized jet streams may imprint an RV on the reference lines themselves. However, its is hard to evaluate how large these effects are when the we integrate along the line of sight, up to the outer atmosphere. The conjugation of three factors point to CO$_2$ features as some of the most stable of the atmosphere, though: 1) the abundance of CO$_2$, 2) the constance of its volume mixing ratio of $\sim$300\,ppmv up to 80\,km, 3) the low absorption line intensities. When taken together, these properties reduce the impact of localized air mass movements on the RV. With the current set of data we can only evaluate the short-term stability of the atmospheric lines. An extended data set with observations under very different atmospheric conditions is now under evaluation. The results will be the subject of a forthcoming paper (Figueira et al. {\it in prep.}).

Temperature and pressure changes are known to be one of the main causes for spectrograph's IP variations. It is important to note that CRIRES is cryogenic and thus stabilized in temperature. At the time of writing, the thermal stability of the most important components of the spectrograph is of $\sim$0.05\,K for the prism, 0.2\,K for the grating and 0.3\,K for the Three-Mirror-Astigmat. As a consequence, the IP remains constant and its impact on RV measurements, by introducing line asymmetries, is reduced. 
Another possible lines profile variation comes from atmospheric phenomena. In what concerns these, and as referred before, the eventual variations are seen both by our reference and target spectra. The drawback is that this shows that typical bisector analysis cannot be used in a straightforward way in stellar spectra, as one must remove the effect of the variation of atmospheric lines' bisector before.
It is then not a surprise that bisector shows no correlation with the RV. Still, the low number of points and the possibility of atmospheric profile variations makes this diagnosis very weak. The data do not provide for a finer analysis, which is not, anyway, within the reach of this paper.

The contribution of photocenter errors was negligible in our error budget but that might be not the case if the seeing becomes very low, as experienced by \cite{2005A&A...431.1105B}. However, our situation is fundamentally different, because our reference lines will be affected by affected by these photocenter movements in the same way than the object lines. In any case, a simple way of reducing these effect is the usage of a fiber entrance, providing for an homogeneous illumination over a wide range of seeing conditions.

The fact that PHOENIX-derived masks led to a lower precision than masks derived from the data themselves is quite insightful on its own. It states that the line's position precision as calculated by state of-the art models is lower than the one that can be measured from high S/N spectra. Some RV determination methods are more sensitive to this fact than others. In particular those that obtain the RV by searching for a match between stellar models and observed spectra struggle against an additional and non-negligible source of noise.

While the application of gas cells on general-purpose slit spectrographs such as CRIRES is very tempting, their yielded precision remains to be assessed. As already discussed by  \cite{2009ApJ...692.1590M}, the usage of the gas cell as a simultaneous reference superimposed on the stellar spectrum is limited by the definition of the stellar continuum. In the case of M stars, one of the main drivers for RV in the IR, the continuum is shaped by a dense forest of unresolved spectral lines. These features are convolved with the atmospheric transmission, which is variable through time. These factors together make RV determinations much more demanding, time and photon consuming and, in many cases, degenerated. In our particular approach, such a poor continuum definition would introduce systematic errors in the absorption lines wavelength assignment, which would propagate to the final calculated RV.

Several infrared spectrographs dedicated to RV measurements are being designed, such as NAHUAL \citep{2005AN....326.1015M} or SPIROU\footnote{$http://www.ast.obs-mip.fr/article.php3?id\_article=637$}. These spectrographs aim at covering a wide spectral range simultaneously, from around 1 to 2.2\,$\mu m$. If a gas cell on the main optical path is chosen as the calibration procedure, the yielded precision will vary locally as a function of stellar continuum definition and atmospheric subtraction quality. On the other hand, using this gas-cell on a twin optical path with an homogeneous continuum source or other methods such as a laser coupled to a Fabry-Perot resonant cavity, allows a simple workaround of these problems. However, as discussed before, this approach is only possible in a fiber-fed, stable spectrograph. 


\section{Conclusions}

 We proved that RV measurements with a precision of 5-10\,m/s are within reach of CRIRES while using carefully selected atmospheric features as a wavelength reference. This precision was obtained on an RV standard, HD\,108309 over a time span of roughly one week. TW\,Hya RV variation in the IR is not compatible with its optical counterpart, showing that a stellar spot is the best explanation for the observed RV variations. This confirms our previously published results \citep{2008A&A...489L...9H} with better precision and consistency. We also showed that the RV variation of a star harboring a planet, Gl\,86, is well reproduced by CRIRES data with a precision which is compatible with photon-noise estimations and at the same noise level. 

\smallskip

The usage of the Earth's atmosphere as a free gas-cell is still to be fully assessed, and the outcome of these studies will, in all probability, shape the design of future dedicated IR spectrographs. In order to do so more lengthy data sets spanning a longer time-span are required. However, while this method seems promising and able to compete with regular gas-cells approaches, it remains to be proven that a higher precision is possible.
Strong limitations might arise below 5 m/s steaming from sources such as the knowledge of the precise wavelength of telluric lines or their stability over long time-scales.

\begin{acknowledgements}
      P.F. would like to thank every researcher who contributed to the enthusiastic, fast-paced and extremely friendly environments of both Observatoire de Gen\`{e}ve and ESO Santiago. This work was shaped by the passionate discussions it triggered. Support from the Funda\c{c}\~{a}o para Ci\^{e}ncia e a Tecnologia (Portugal) to P. F. in the form of a scholarship (reference SFRH/BD/21502/2005) is gratefully acknowledged. NCS would like to thank the support by the European Research Council/European Community under the FP7 through a Starting Grant, as well from Funda\c{c}\~ao para a Ci\^encia e a Tecnologia (FCT), Portugal, through program Ci\^encia\,2007, and in the form of grants reference PTDC/CTE-AST/098528/2008 and PTDC/CTE-AST/098604/2008.
 We thank the ESO Director Office for granting DDT observations. We also thank the anonymous referee for his/her valuable suggestions on how to improve the manuscript.

\end{acknowledgements}

\bibliographystyle{aa} 
\bibliography{Mybibliog} 

\end{document}